# A Framework for Optimal Matching for Causal Inference

Nathan Kallus
Cornell University and Cornell Tech
kallus@cornell.edu

## Abstract

We propose a novel framework for matching estimators for causal effect from observational data that is based on minimizing the dual norm of estimation error when expressed as an operator. We show that many popular matching estimators can be expressed as optimal in this framework, including nearest-neighbor matching, coarsened exact matching, and mean-matched sampling. This reveals their motivation and aptness as structural priors formulated by embedding the effect in a particular functional space. This also gives rise to a range of new, kernel-based matching estimators that arise when one embeds the effect in a reproducing kernel Hilbert space. Depending on the case, these estimators can be found using either quadratic optimization or integer optimization. We show that estimators based on universal kernels are universally consistent without model specification. In empirical results using both synthetic and real data, the new, kernel-based estimators outperform all standard causal estimators in estimation error.

## 1 Introduction

Compared to controlled experiments, observational studies are uniquely characterized by a lack of control on membership in the treatment and control groups. While in controlled experimentation, randomization ensures comparability and hence unbiased and consistent estimation of effect; in observational studies, valid inference about a causal effect of treatment requires adjusting the groups so that they become comparable. Comparable for the purpose of causal inference means as similar as possible in some observed covariates. The covariates constitute the relevant information known about each observational subject and, as long as these covariates account for any confounding between the effects of treatment and the effects of self-selection, making the groups comparable with respect to these makes the groups comparable for the purpose of causal inference.

Matching has been some of the most popular ways to achieve this comparability [7, 22, 32]. In matching, we sample a subset from the groups to get samples that are more similar to one another than the original samples. More generally, we may re-weight the original sample, where weights that are integer multiples correspond to (multi)subsets. For example, in nearest-neighbor matching (NNM) [21], one composes a matched sample out of pairs of treated and control subjects so that the total pairwise distance between covariate vectors is small or even minimal, mimicking a randomized matched-pair experiment [12]. If we allow subjects to be paired with replacement, we can have a sample with duplicates, resulting in weights that correspond to a multisubset rather than a regular subset. In coarsened exact matching (CEM) [15], one coarsens the covariates to create strata and re-weights the samples so that they have equal frequency in each stratum, mimicking a randomized block experiment [8], which results in general weights that may not correspond to taking a subset of the data.

We focus on these and similar matching estimators that balance the covariates themselves rather than imputed propensity scores, as in propensity score matching (PSM) [23]. Matching on covariates addresses imbalance and not just confounding [18, 17]. Nonetheless, we include PSM in numerical experiments.

In this paper, we develop a novel and encompassing framework for estimators that balance the covariates via matching (in the broader re-weighting sense). There are many different such estimators and each addresses imbalance differently. Our framework teases out how a particular notion of imbalance corresponds to a notion of structure. By decomposing the error

A Framework for Optimal Matching for Causal Inference

of matching estimators, we formulate the error of the estimator as an operator on the conditional expectation function of outcomes given covariates. This conditional expectation function is unknown (or else there would be no need to conduct the study) and when one considers what the *worst-case* error may be over a space of possible such functions one recovers the *dual norm of the error* if the space is a Banach space. The dual norm of the error is an observable quantity, expressed only in terms of the given data. We term any estimator that chooses matched subsamples by minimizing this quantity as *error-dual-norm minimizing* (EDNM). A surprising result is that a great variety of standard methods used in the practice of causal inference are all EDNM. This observation leads us to consider new methods that are EDNM. Using reproducing kernel Hilbert spaces (RKHS) to express structure we obtain a new class of kernel-based matching estimators for causal effects. These have desirable properties like consistency and perform exceptionally well in practice. All proofs are given in the supplement.

**Set Up.** We begin by describing the set up. We consider an observational study with $n$ subjects, indexed $i = 1, \ldots, n$. We let this order be arbitrary so that the subjects are exchangeable (later, we consider subjects comprising an iid process). Of these, $n_1$ received a treatment whose effect is of interest (denoted by $T_i = 1$) and $n_0$ received a control treatment against which we want to compare (denoted by $T_i = 0$). Let $\mathcal{T}_0 = \{i : T_i = 0\}$ and $\mathcal{T}_1 = \{i : T_i = 1\}$ be the sets of subjects that received treatment and control, respectively. We let $T = (T_1, \ldots, T_n)$.

Using Neyman-Rubin potential outcome notation [29], we let $Y_i(0), Y_i(1)$ be the (real-valued) potential outcomes for subject $i$. We observe the outcome for the treatment to which subject $i$ was exposed, $Y_i = Y_i(T_i)$. And, $Y(1 - T_i)$ represents the unobserved, counterfactual outcome we would have observed if subject $i$ were exposed to the opposite treatment. $Y(1 - T_i)$ is *missing data*. Throughout the paper, for these to be well defined, we assume that the stable unit treatment value assumption (SUTVA) holds [26].

Let $X_i$, taking values in some $\mathcal{X}$, be the side covariates that we observe for subject $i$. Let $X = (X_1, \ldots, X_n)$ denote the collection of all baseline covariates of all $n$ subjects, which constitues part of the observed data. The space $\mathcal{X}$ is general; assumptions about it will be specified as necessary. As an example, it can be composed real-valued vectors $\mathcal{X} \subseteq \mathbb{R}^d$ that include both discrete (dummy) and continuous variables.

We denote by $\text{TE}_i = Y_i(1) - Y_i(0)$ the unobservable causal treatment effect for subject $i$. The primary quantity of interest for estimation is the *sample average (causal) treatment effect on the treated sample*:
$\text{SATT} = \frac{1}{n_1} \sum_{i \in \mathcal{T}_1} \text{TE}_i = \frac{1}{n_1} \sum_{i=1}^n T_i(Y_i(1) - Y_i(0))$.
We consider estimators for SATT based on matching in the form of *re-weighting*. We restrict to *honest* weights that only depend on the observed $X$, $T$ and not on any observed outcome data. (If we used outcome data one might complain that we are mining for an effect that is not there.) In particular, we consider the choice of a function $W = W(X, T)$ that produces a weight $W_i \in \mathbb{R}_+$ for each subject $i$, leading to the estimator

$$\hat{\tau}_W = \sum_{i=1}^n (-1)^{T_i+1} W_i Y_i.$$

Because we are estimating SATT and we in fact observe $Y_i(1)$ for each $i \in \mathcal{T}_1$, we always set $W_i = 1/n_1$ for $i \in \mathcal{T}_1$, leading to estimators of the form

$$\hat{\tau}_W = \tfrac{1}{n_1} \sum_{i \in \mathcal{T}_1} Y_i - \sum_{i \in \mathcal{T}_0} W_i Y_i.$$

We also always assume $\sum_{i \in \mathcal{T}_0} W_i = 1$.

We let $\mathcal{W} = \mathcal{W}_0 \times \mathcal{W}_1$ denote the space of allowable weights, where $\mathcal{W}_0$ and $\mathcal{W}_1$ are the space of weights for the control and treated sample, respectively. We required that $\mathcal{W}_0 \subseteq \{W_{\mathcal{T}_0} \in \mathbb{R}^{\mathcal{T}_0} : \sum_{i \in \mathcal{T}_0} W_i = 1\}$ and that $\mathcal{W}_1 = \{(1/n_1, \ldots, 1/n_1)\}$. If all weights in $\mathcal{W}_0$ are rational with a fixed denominator, then $\hat{\tau}_W$ corresponds to constructing a (multi-)set from the control subjects to match the treated sample. We note some special cases of $\mathcal{W}_0$ that correspond to a variety of existing classes of estimators for SATT:
- Probability (convex combination) weights:
$\mathcal{W}_0^{\text{probability}} = \left\{W_{\mathcal{T}_0} \in \mathbb{R}_+^{\mathcal{T}_0} : \sum_{i \in \mathcal{T}_0} W_i = 1\right\}$.
- Bounded probability weights:
$\mathcal{W}_0^{b\text{-bounded}} = \mathcal{W}_0^{\text{probability}} \cap [0, b]^{n_0}$.
- Subsets of cardinality $n'_0$ (without replacement):
$\mathcal{W}_0^{n'_0\text{-subset}} = \mathcal{W}_0^{\text{probability}} \cap \{0, 1/n'_0\}^{\mathcal{T}_0}$.
- Subsets of cardinality at least $\underline{n}'_0$:
$\mathcal{W}_0^{\geq \underline{n}'_0\text{-subset}} = \bigcup_{n'_0 = \underline{n}'_0}^{n_0} \mathcal{W}_0^{n'_0\text{-subset}}$.
- Multisubsets of cardinality $n'_0$ (with replacement):
$\mathcal{W}_0^{n'_0\text{-multisubset}} = \mathcal{W}_0^{\text{probability}} \cap \{0, 1/n'_0, 2/n'_0, \ldots\}^{\mathcal{T}_0}$.

A standing assumption in this paper, essential for causal inference from observational data, is that of *weak ignorability in expectation*.

**Assumption 1.** For each $t = 0, 1$ and $i = 1, \ldots, n$, conditioned on $X_i$, $Y_i(t)$ is mean-independent of $T_i$ and each value of $T_i$ is possible. That is, for each $t = 0, 1$ and $i = 1, \ldots, n$,

$$\mathbb{E}\left[Y_i(t) \mid T_i, X_i\right] = \mathbb{E}\left[Y_i(t) \mid X_i\right], \quad \mathbb{P}\left(T_i = t \mid X_i\right) > 0.$$

Ignorability, also known as unconfoundedness, means that we have the right covariates needed to separate the effect of the treatment itself from the effect of self-selection [23]. The form of ignorability we use is termed "weak" because it need only apply for each $t = 0, 1$ separately, and it is termed "in expecta-tion" because only mean-independence, rather than full stochastic independence, is assumed.

## 2 EDNM

**Decomposing the Error.** Denote conditional expectation of the control potential outcome given the covariates $x$ by $f_0(x) = \mathbb{E}\left[Y_i(0)\big|X_i = x\right]$. The non-random function $f_0$ does not depend on $i$ due to exchangeability. By iterated expectation, the residual $\epsilon_i = Y_i(0) - f_0(X_i)$ has mean 0, is mean-independent of $X_i$, and is uncorrelated with any function of $X_i$.

By conditioning on $X_i$, we can decompose the error of the estimator into two terms: error that can be controlled by matching on $X_i$ and the orthogonal residual error, which cannot be controlled by $X_i$ but which disappears in expectation due to ignorability.

**Theorem 1.** $\hat{\tau}_W - \text{SATT} = F_W + E_W$, where

$$F_W = \mathcal{E}(W; f_0), \quad E_W = \frac{1}{n_1}\sum_{i \in \mathcal{T}_1} \epsilon_i - \sum_{i \in \mathcal{T}_0} W_i \epsilon_i,$$

$$\mathcal{E}(W; f) = \frac{1}{n_1}\sum_{i \in \mathcal{T}_1} f(X_i) - \sum_{i \in \mathcal{T}_0} W_i f(X_i).$$

*Moreover, under Assumption 1,*

$$\mathbb{E}\left[\hat{\tau}_W - \text{SATT} \mid X, T\right] = \mathcal{E}(W; f_0), \quad \mathbb{E}\left[E_W \mid X, T\right] = 0.$$

**The Dual Norm of the Error.** The target of matching for causal inference is to eliminate error in comparing the treatment and control samples. Theorem 1 provides an explicit form of the controllable error in terms of the observed covariates $X$. However, it involves the unknown function $f_0 : \mathcal{X} \to \mathbb{R}$. As alluded to in Sec. 1, we consider matching schemes that guard against any possible such function by minimizing the worst-case error over the unit ball of a Banach space. A normed vector space is a Banach space if the corresponding metric space is complete (see [19], and [24], Ch. 10 for more on Banach spaces).

Let $\mathcal{V}$ denote the vector space of all functions $\mathcal{X} \to \mathbb{R}$ under usual pointwise addition and scaling. Let $\mathcal{F} \subseteq \mathcal{V}$ be a subspace of functions, against which we wish to guard. Endow this space with a semi-norm $\|\cdot\| : \mathcal{F} \to \mathbb{R}$ (a semi-norm can assign zero magnitude to nonzero vectors). For $f \notin \mathcal{F}$, let us write $\|f\| = \infty$. Thus, the assumption that $f_0 \in \mathcal{F}$ is encapsulated by $\|f_0\| < \infty$.

Given only that $\|f_0\| < \infty$, we will consider matching schemes that choose $W$ to minimize the worst-case error, $\max_{\|f\| \leq \|f_0\|} |\mathcal{E}(W; f)| = \|f_0\| \max_{\|f\| \leq 1} \mathcal{E}(W; f)$, where the equality holds because $\mathcal{E}(W; \alpha f) = \alpha \mathcal{E}(W; f)$ is degree-1 homogeneous and $\|\alpha f\| = |\alpha| \|f\|$ is degree-1 positively homogeneous and symmetric. Clearly, it only matters that $\|f_0\| < \infty$ and the particular finite value of it does not change which $W$ minimizes the above. In light of this, we define the *worst-case error* as

$$\mathfrak{E}(W; \mathcal{F}) = \max_{\|f\| \leq 1} \mathcal{E}(W; f).$$

Since $\sum_{i=1}^{n}(-1)^{T_i+1}W_i = 0$, we have that $\mathcal{E}(W; f)$ is invariant to constant shifts to $f$, i.e., $\mathcal{E}(W; f) = \mathcal{E}(W; f+c)$, where $c \in \mathbb{R}$ represents a constant function $x \mapsto c$. To eliminate this irrelevant mode of $\mathcal{F}$, we can just consider the quotient space $\mathcal{F}/\mathbb{R}$, which consists of the equivalence classes $[f] = \{f + c : c \in \mathbb{R}\}$ endowed with the norm $\|[f]\| = \min_{c \in \mathbb{R}} \|f + c\|$. Note that by construction, $\mathcal{E}(W; [f]) = \mathcal{E}(W, f)$ is well defined. For brevity, we will simply refer to $\mathcal{F}$ and $\|\cdot\|$ when we mean $\mathcal{F}/\mathbb{R}$ and the corresponding norm.

We assume $(\mathcal{F}, \|\cdot\|)$ satisfies the following conditions:

**Assumption 2.** *The space $\mathcal{F}$ is a Banach space.*

**Assumption 3.** *For each $W \in \mathcal{W}$, $f \mapsto \mathcal{E}(W; f)$ is a continuous mapping $\mathcal{F} \to \mathbb{R}$.*

Since $\mathcal{E}(W, f)$ is also linear in $f$, these assumptions imply that, for each $W$, the operator $\mathcal{E}(W, \cdot)$ is in the continuous dual space of $\mathcal{F}$. Hence,

$$\mathfrak{E}(W; \mathcal{F}) = \|\mathcal{E}(W; \cdot)\|_*$$

is *precisely* the dual norm of the error, where the dual norm of a continuous linear operator $A$ on a Banach space with norm $\|\cdot\|$ is $\|A\|_* = \sup_{\|u\| \leq 1} A(u)$. This also guarantees that $\mathfrak{E}(W; \mathcal{F})$ is finite and well-defined.

**Definition 1.** *A matching method $W(T, X)$ is said to be error-dual-nrom (EDNM) if for some $\mathcal{W}$ and $(\mathcal{F}, \|\cdot\|)$ satisfying Assumptions 2 and 3 we have*

$$W(T, X) \in \operatorname{argmin}_{W \in \mathcal{W}} \mathfrak{E}(W; \mathcal{F}) \neq \mathcal{W}.$$

Let $\mathfrak{E}_{\min}(\mathcal{F}) = \min_{W \in \mathcal{W}} \mathfrak{E}(W; \mathcal{F})$ be the optimal value. Clearly, if a matching method $W(T, X)$ is EDNM with $(\mathcal{F}, \|\cdot\|)$ and $\mathcal{W}$ then the error of $\hat{\tau}_W$ is bounded by $|\mathcal{E}(W; f_0)| \leq \|f_0\| \mathfrak{E}_{\min}(\mathcal{F})$.

### 2.1 Existing Methods as EDNM

Surprisingly, many methods for causal inference that are standard in practice are also in fact EDNM. On the one hand, this interpretation gets at the core of the structural motivations behind many of these methods (e.g., "if you believe the conditional expectation is Lipschitz and nothing more then you should pairwise match") and allows one to choose a method appropriate to one's beliefs about problem structure. On the other hand, these results provide motivation that EDNM is the *right* framework in which to think about matching for causal inference and this motivates us to consider new EDNM methods in Sec. 2.2.

**Nearest-Neighbor Matching.** NNM is by far the most common matching method. In NNM, each treated subject is paired with one control subject so that the sum of pairwise distances is minimized as measured by some distance metric $\delta(x, x')$ on $\mathcal{X}$ [21]. Usually, the Mahalanobis metric is used: $\delta^2(x, x') =$

$(x - x')\hat{\Sigma}^{-1}(x - x')$, where $\hat{\Sigma}$ is the pooled sample covariance matrix. NNM can be done either without replacement (each control subject used at most once; aka one-to-one) or with replacement (control subjects may be reused; aka many-to-one). The estimate of SATT is the average pairwise differences of outcomes. This estimator is exactly $\hat{\tau}_W$ where the weight on control subject $i$ is $1/n_1$ times the number of times subject $i$ was matched, i.e., the matched control sample is the (multi-)set of control subjects that got matched to treated subjects. NNM is EDNM as we show next.

**Theorem 2.** *NNM with distance metric $\delta(x, x')$ with replacement and without replacement are both EDNM with $\|f\| = \sup_{x \neq x'} \frac{f(x) - f(x')}{\delta(x,x')}$, the Lipschitz constant of $f$; $\mathcal{F} = \{f : \|f\| < \infty\}$; $\mathcal{W}_0$ is either $\mathcal{W}_0^{probability}$ or $\mathcal{W}_0^{n_1\text{-}multisubset}$ if with replacement; and $\mathcal{W}_0$ is either $\mathcal{W}_0^{n_1^{-1}\text{-}bounded}$ or $\mathcal{W}_0^{n_1\text{-}subset}$ if without replacement.*

Note that even if the weights are not restricted to be multiples of $1/n_1$, the *optimal* unrestricted weights will end up to be multiples of $1/n_1$ regardless. That is, the optimal general-form weighting is optimal subset matching for Lipschitz functions.

Note that $(\mathcal{F}, \|\cdot\|)$ is *not* a Banach space. In particular, constant functions have zero Lipschitz constant. However, as required, $\mathcal{F}/\mathbb{R}$ is a Banach space and evaluation differences are continuous because they are bounded by the magnitude.

Algorithmically, NNM with replacement amounts to finding the control subject of minimal distance to each treated subject in a greedy manner. NNM without replacement amounts to minimum-sum-of-distances bipartite matching with unbalanced parts, which is easily solved by the Ford-Fulkerson algorithm [9].

A close cousin is caliper matching whereby we only match subjects that are within a distance $\delta_0$ from one another. This method is also EDNM.

**Corollary 3.** *Caliper matching with pairwise distance metric $\delta(x, x')$ and threshold $\delta_0$ with replacement and without replacement are both EDNM (if feasible) with $\|f\| = \sup_{x \neq x'} \frac{f(x) - f(x')}{\max\{\delta_0, \delta(x,x')\}}$; $\mathcal{F} = \{f : \|f\| < \infty\}$; $\mathcal{W}_0$ is either $\mathcal{W}_0^{probability}$ or $\mathcal{W}_0^{n_1\text{-}multisubset}$ if with replacement; and $\mathcal{W}_0$ is either $\mathcal{W}_0^{n_1^{-1}\text{-}bounded}$ or $\mathcal{W}_0^{n_1\text{-}subset}$ if without replacement.*

**Coarsened Exact Matching.** CEM [15] is a matching method whereby one coarsens the covariates into a few ($M$) strata via a coarsening function $C : \mathcal{X} \to \{1, \ldots, M\}$, and then matches exactly within each stratum. For example, if there are 5 treated subjects and 3 control subjects in a given stratum then each of the 3 control subjects is given weight proportional to 5/3, whereas if there were 0 treated subject the weights would be 0. The case of a stratum containing only treated subjects is not allowed (no extrapolation).([16] suggests that in this case one not estimate SATT.) Under Assumption 1, this case happens with vanishing probability. CEM is also EDNM.

**Theorem 4.** *CEM with coarsening function $C : \mathcal{X} \to \{1, \ldots, M\}$ is EDNM with $\mathcal{F} = \{f : |f^{-1}(C^{-1}(j))| = 1 \ \forall j = 1, \ldots, M\}$ (i.e., piece-wise constant on partitions); $\|f\| = \sup_{x \in \mathcal{X}} |f(x)|$ for $f \in \mathcal{F}$, otherwise $\infty$; and $\mathcal{W}_0$ is $\mathcal{W}_0^{probability}$; assuming no extrapolation.*

**Mean-Matched Sampling.** Often, practitioners evaluate the quality of a matched control sample by measuring the Mahalanobis distance between the matched control sample and the treated sample: $M_V(W) = \|V^{1/2}(\frac{1}{n_1} \sum_{i \in \mathcal{T}_1} X_i - \sum_{i \in \mathcal{T}_0} W_i X_i)\|_2$, where $\mathcal{X} \subseteq \mathbb{R}^d$ and $V$ is some positive semidefinite matrix usually taken to be $V = \hat{\Sigma}_0^\dagger$, the inverse sample covariance matrix of $X|T=0$. This distance is a rotated 2-norm between the sample means. Mean-matched sampling are methods that find match a control sample of prescribed size to reduce this distance [10, 25] and optimal mean-matching sampling (OMMS) fully minimizes this distance and is EDNM.

**Theorem 5.** *OMMS of $n_0'$ subjects from the control sample (with or without replacement) is EDNM with $\mathcal{F} = \{x \mapsto \beta_0 + \beta^T x : \beta \in \mathbb{R}^d\}$; $\|x \mapsto \beta_0 + \beta^T x\| = \sqrt{\beta^T V \beta + \beta_0^2}$ and $\|f\| = \infty$ otherwise; and $\mathcal{W}_0$ is $\mathcal{W}_0^{n_0'\text{-}multisubset}$ if with replacement or $\mathcal{W}_0^{n_0'\text{-}subset}$ if without, respectively.*

Since finite, the space $(\mathcal{F}, \|\cdot\|)$ is always a Banach space and evaluations (and hence their differences) are always continuous. See Thms. 5.33 and 5.35 of [14].

## 2.2 Kernel Matching

In the previous section we saw that a variety of standard methods for causal inference are EDNM. Each was recovered using a different form of structure on the conditional expectations of outcomes. In this section we develop a range of new EDNM based on kernels and their corresponding reproducing kernel Hilbert spaces (RKHS). Kernels are standard in machine learning (ML) as ways to generalize the structure of learned conditional expectation functions, like classifiers or regressors [27]. Kernels also have many applications in statistics such as in independence testing [3, 11, 34] and goodness-of-fit testing [11]. The same way kernels are used to generalize the structure of learned functions in ML, we can use these to generalize the structure of $f_0$. This will lead to new methods for causal inference that are potentially very powerful.

A Hilbert space is an inner-product space such that the norm induced by the inner product, $\|f\|^2 = \langle f, f \rangle$, yields a Banach space. An RKHS $\mathcal{F}$ is a Hilbert space of functions for which, for every $x \in \mathcal{X}$, the map $f \mapsto f(x)$ is a continuous mapping [3]. Continuity and the Riesz representation theorem imply that for each $x \in \mathcal{X}$ there is $\mathcal{K}(x, \cdot) \in \mathcal{F}$ such that $\langle \mathcal{K}(x, \cdot), f(\cdot) \rangle = f(x)$ for every $f \in \mathcal{F}$. The symmetric map $\mathcal{K} : \mathcal{X} \times \mathcal{X} \to \mathbb{R}$ is called the reproducing kernel of $\mathcal{F}$. The name is motivated by the fact that $\mathcal{F} = \text{closure}(\text{span}\{\mathcal{K}(x, \cdot) : x \in \mathcal{X}\})$. Thus $\mathcal{K}$ fully characterizes $\mathcal{F}$. Prominent examples of kernels for $\mathcal{X} \subset \mathbb{R}^d$ are: the polynomial kernel $\mathcal{K}_{\sigma,p}(x, x') = (1 + x^T x'/(\sigma^2 p))^p$, whose RKHS spans the finite-dimensional space of all polynomials of degree up to $p$; the exponential kernel $\mathcal{K}_\sigma(x, x') = e^{x^T x'/\sigma^2}$, the infinite-dimensional limit of the polynomial kernel; and the Gaussian kernel $\mathcal{K}_\sigma(x, x') = e^{-\|x-x'\|^2/\sigma^2}$. The corresponding RKHS is infinite-dimensional [31]. For $X \in \mathcal{X}^n$ and a kernel $\mathcal{K}$, the Gram matrix is $K_{ij} = \mathcal{K}(X_i, X_j)$, which is always positive semi-definite (PSD). Generally, one would normalize the covariate data before putting it in a kernel using an affine transformation so that the control sample has zero sample mean and identity sample covariance.

Some kernels have a special property, known as universality, that allows them to approximate any continuous function arbitrarily well. Both the Gaussian and exponential kernels are universal [30].

**Definition 2.** For $\mathcal{X}$ compact Hausdorff, a kernel is *universal* if for any continuous function $g : \mathcal{X} \to \mathbb{R}$ and $\epsilon > 0$, there exists $f \in \mathcal{F}$ in the corresponding RKHS such that $\sup_{x \in \mathcal{X}} |f(x) - g(x)| \leq \epsilon$.

Note that any RKHS $\mathcal{F}$ satisfies Assumptions 2 and 3. As such it gives rise to EDNM matching methods.

**Theorem 6.** *Let $\mathcal{F}$ be an RKHS with kernel $\mathcal{K}$. Let $K$ be the Gram matrix on $X$. Then,*

$$\mathfrak{E}(W; \mathcal{F}) = (\tfrac{1}{n_1^2} e_{n_1}^T K_{\mathcal{T}_1, \mathcal{T}_1} e_{n_1} + W_{\mathcal{T}_0}^T K_{\mathcal{T}_0 \mathcal{T}_0} W_{\mathcal{T}_0}$$
$$- \tfrac{2}{n_1} e_{n_1}^T K_{\mathcal{T}_1, \mathcal{T}_0} W_{\mathcal{T}_0})^{1/2}.$$

The above theorem makes clear that kernel matching with a linear kernel $\mathcal{K}_V(x, x') = x^T V x'$ is exactly equivalent to mean-matched sampling, i.e., it leads to $\mathfrak{E}(W; \mathcal{F}) = M_V(W)$. Moreover, if $W_{\mathcal{T}_0} \in \{0, 1/n_0'\}^{\mathcal{T}_0}$ then $\mathfrak{E}(W; \mathcal{F})$ is exactly the kernel *maximum mean discrepancy* (MMD) statistic between the treated sample and the matched control sample. Kernel MMD is a common test statistic in two-sample goodness-of-fit testing [11, 28]. We can interpret minimizing this discrepancy as trying to make the two samples appear to come from the exact same distribution.

Next, we review the various possible methods this can give rise to. In the following, we let $k_0 = K_{\mathcal{T}_0 \mathcal{T}_1} e_{n_1}/n_1$.

**Kernel Matching with Probability Weights.** For probability weights, we can formulate a linearly-constrained convex-quadratic optimization problem to find the optimal weights:

$$\text{argmin}_{W_{\mathcal{T}_0} \in \mathcal{W}_0^{\text{probability}}} \mathfrak{E}(W; \mathcal{F})$$
$$= \text{argmin}_{W \in \mathbb{R}_+^{n_0} : e_{n_0}^T W = 1} \left( W K_{\mathcal{T}_0 \mathcal{T}_0} W - 2 k_0^T W \right).$$

This problem can be solved in polynomial time with interior point methods [4] and is amenable to solution with off-the-shelf solvers like Gurobi.

**Kernel Multisubset Matching.** For matching with replacement, we can formulate a linear-integer-constrained convex-quadratic optimization problem to find the optimal weights:

$$\text{argmin}_{W_{\mathcal{T}_0} \in \mathcal{W}_0^{n_0'\text{-multisubset}}} \mathfrak{E}(W; \mathcal{F})$$
$$= \tfrac{1}{n_0'} \text{argmin}_{W' \in \mathbb{Z}^{n_0} : e_{n_0}^T W' = n_0'} \left( \tfrac{1}{n_0'} W' K_{\mathcal{T}_0 \mathcal{T}_0} W' - 2 k_0^T W' \right),$$

where we used the change of variables $W' = n_0' W_{\mathcal{T}_0}$. This problem is NP-hard (reducible to number partitioning for $\text{rank}(K_{\mathcal{T}_0 \mathcal{T}_0}) = 1$), but it is also amenable to solution by off-the-shelf integer programming solvers like Gurobi.

**Kernel Subset Matching.** For matching without replacement, we can formulate a linear-integer-constrainted convex-quadratic optimization problem to find the optimal weights:

$$\text{argmin}_{W_{\mathcal{T}_0} \in \mathcal{W}_0^{n_0'\text{-subset}}} \mathfrak{E}(W; \mathcal{F})$$
$$= \tfrac{1}{n_0'} \text{argmin}_{W' \in \{0,1\}^{n_0} : e_{n_0}^T W' = n_0'} \left( \tfrac{1}{n_0'} W' K_{\mathcal{T}_0 \mathcal{T}_0} W' - 2 k_0^T W' \right).$$

Again, the problem is generally "hard" but can be solved in practice using off-the-shelf integer programming solvers.

### 2.3 Consistency

Next, we express conditions for EDNM estimators to have error converging to zero.

**Definition 3.** A Banach space is said to be *B-convex* if there exists $N \in \mathbb{N}$ and $\eta < N$ such that for every $g_1, \ldots, g_N$ with $\|g_i\| \leq 1 \,\forall i$ there exists a choice of signs so that $\|\pm g_1 \pm \cdots \pm g_N\| \leq \eta$.

It is easy to verify that all the Banach spaces so far considered are $B$-convex. All Hilbert spaces and all finite-dimensional Banach spaces are $B$-convex [19, Ch. 9].

**Theorem 7.** *Suppose Assumption 1 holds and that (1) the subjects $i = 1, 2, \ldots$ form an independent and identically distributed process; (2) for each $n$, $W$ is EDNM with $(\mathcal{F}, \|\cdot\|)$, $\mathcal{W}$; (3) $\mathcal{W}^{\geq n_0'\text{-subset}} \subset \mathcal{W}_0$ for some fixed $\underline{n}_0' \geq 1$; (4) $f_0 \in \text{Closure}_\infty(\mathcal{F})$, i.e., $\forall \epsilon > 0, \exists g_0 \in \mathcal{F} : \sup_{x \in \mathcal{X}} |f_0(x) - g_0(x)| \leq \epsilon$; and (5)* $\mathbb{E}[\max_{\|f\|\leq 1} |f(X_1) - \mathbb{E}[f(X_1)|T_1 = 1]|^\nu | T_1 = 1] < \infty$ and either (a) $\mathcal{F}$ is B-convex and $\nu = 2$ or (b) $\mathcal{F}$ is a Hilbert space and $\nu = 1$. Then, $F_W \to 0$ almost surely.

Note that if, in addition, $\|W\| \to 0$ (e.g., as in subset matching with $n'_0 \to \infty$) then $E_W \to 0$ and hence, by Theorem 1, $\hat{\tau}_W - \text{SATT} \to 0$.

Clearly, one way to satisfy condition (4) is to have $f_0 \in \mathcal{F}$, i.e., to make the correct structural assumption. But, it is sufficient that $f_0$ is close to $\mathcal{F}$. Both universal RKHSs (for kernel matching) and the space of Lipschitz functions (for NNM) are dense in continuous functions (i.e., they reside in their closure) in the sense of condition (4).

For kernel matching with kernel $\mathcal{K}$, to satisfy condition (5), it is sufficient that $\mathbb{E}[\sqrt{\mathcal{K}(X_1, X_1)} \mid T_1 = 1] < \infty$.

## 3 Empirical Results

In this section, we study empirically the comparative efficiency of various causal estimators, including our new kernel estimators. First, we consider a simple synthetic observational study that allows us to investigate the interaction between underlying structure and matching method used. Second, we consider an observational study based on a dataset compiled by [13] from the Infant Health and Development Program [5].

**Fictitious Study.** Consider the following fictitious observational study with one treatment and control. Subjects are drawn at random from a population. For each subject we observe a two-dimensional vector of covariates $X_i \in \mathbb{R}^2$. In the population, these are distributed as uniform on $[-1, 1]^2$. Each subject has either received treatment or control and we observe $T_i$. In the population, $T_i$ is distributed as Bernoulli with probability $0.8/\left(1 + \sqrt{2}\|X_i\|_2\right)$, which ranges $0.27 \sim 0.8$.

The potential outcomes are $Y_i(0) = f_0(X_i) + \epsilon_{0i}$, $Y_i(1) = f_1(X_i) + \epsilon_{1i}$, where $\epsilon_{0i}, \epsilon_{1i} \sim \mathcal{N}(0, 0.1)$ is independent noise. We focus on the case of small residual noise (variance not explained by $X_i$) so to tease out the comparative efficiency in matching $X$ (if residual noise is big, any method that only matches on $X$ will do badly). We let $f_1$ be any function whatsoever. We consider a variety of possible cases for $f_0$:
- $\ell_1$ norm: $f_0(x) = |x_1| + |x_2|$;
- quadratic: $f_0(x) = (x_1 + x_2) + (x_1 + x_2)^2$;
- cubic: $f_0(x) = (x_1 + x_2)^2 + (x_1 + x_2)^3$;
- sinusoidal: $f_0(x) = \sin(\pi(x_1 + x_2)) + \cos(\pi(x_1 - x_2))$.

For each $n = 10, 20, \ldots, 300$, we produce 100 replicates. For each, we consider a variety of estimators:
- No matching: we take the whole control sample to be the matched sample ($W_i = 1/n_0$);
- One-to-one: we match $n_1$ control subjects using NNM without replacement, i.e., using optimal bipartite matching on the matrix of pairwise Mahalanobis distances between treated and control subjects;
- CEM: we find the largest $b \geq 0$ such that coarsening each of the covariates into even bins $\{[-1, -1 + 2^{b-1}), \ldots, [1 - 2^{b-1}, 1]\}$ leaves no box (product of two bins) that contains only treated subjects, then we perform exact matching within each box;
- Mahal. means: we match $n_1$ control subjects with replacement to minimize the Mahalanobis distance between the means of the two samples (same as matching $n_1$ control subjects with replacement using kernel matching with the linear kernel);
- PSM: we match $n_1$ control subjects using propensity score matching by fitting a logistic regression to impute propensity scores and doing optimal bipartite matching on imputed scores;
- Quad kernel weight: we use kernel matching with probability weights and the quadratic kernel;
- Exp kernel weight: we use kernel matching with probability weights and the exponential kernel;
- Gauss kernel weight: we use kernel matching with probability weights and the Gaussian kernel;
- Exp kernel match: we match $n_1$ control subjects with replacement using kernel matching with the exponential kernel; and
- Gauss kernel match: we match $n_1$ control subjects with replacement using kernel matching with the Gaussian kernel.

We let $\sigma = 1$ for all kernels. We use Gurobi v6.5 (www.gurobi.com) to solve all quadratic and integer optimization problems. For each estimator, we compute $\hat{\tau}_W - \text{SATT}$. Then, we measure the RMSE over the 100 replicates, $\text{RMSE} = (\hat{\mathbb{E}}_{100} \left[ (\hat{\tau}_W - \text{SATT})^2 \right])^{1/2}$. We plot the results in Figs. 1(a-d). Note the log scale.

The results clearly show the power of our approach. In each case, every one of our exponential- or Gaussian-kernel-based estimators outperforms standard causal estimators by an order of magnitude (base 10). The advantage is particularly noticeable in smaller samples (notice the initial sharp drop in most plots). Indeed, it can be difficult to find a good control pair for every treated subject in small samples, and similarly it can be difficult to have a fine enough coarsening of the data without creating a stratum that only has treated subjects. Whereas, at the same time, by *optimizing* the mismatch as characterized by the dual norm of the error one can achieve small mismatch with even small samples (in agreement with the observation made by [17] about multi-objective partitioning).

Another observation is that matching based on parametric models can be fragile. This can be seen here for PSM, which is based on a misspecified logistic model, and also for estimators that match on $X$ itself. We also see that mean-matched sampling does very poorly in every example, even doing worse than no matching. Indeed, matching the means only makes sense if the effect is *purely linear*. A linear model assumption is very fragile and even small violations can trip up mean-matched sampling. Similarly, matching per the quadratic kernel depends on an assumption of quadratic effect. Indeed, the estimator based on the quadratic kernel does the best of all estimators when the effect is quadratic (panel b). However, unlike linear, a quadratic model is generally more robust as quadratics can better approximate a wider range of functions. Accordingly, we see that the estimator based on the quadratic kernel has reasonable performance even when the effect is not quadratic (panels a and c), while extreme violations trip it up (panel d).

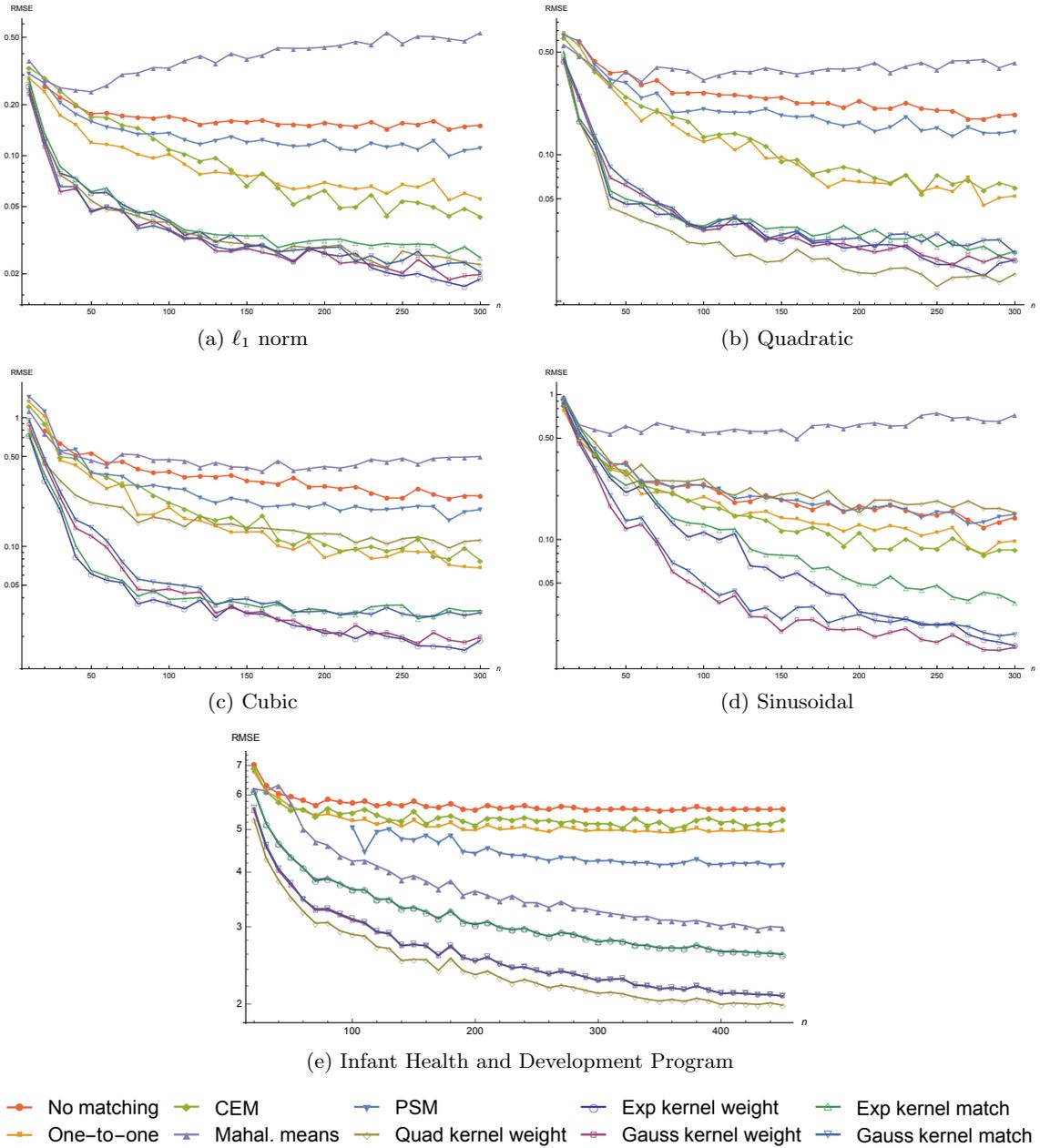

Figure 1: RMSE (log scale) of various causal estimators for various effect functions in Section 3

Overall, the universal kernels (exponential and Gaussian) seem to do the best by far. They appear to provide a good balance between generality of model with efficiency of balancing. They are general enough so that we can ensure consistency even if the true effect is not in the corresponding RKHS. And, fully optimizing mismatch as measured by the dual norm of the error in their RKHS can lead to small objective value even for moderate $n$.

**Infant Health and Development Program (IHDP).** IHDP was a randomized experiment intended to measure the effect of a program consisting of child care and home visits from a trained provider on early child development [5], as measured through cognitive test scores. The data form this study was used by [13] to evaluate causal estimators where each study subject is one child. We use a similar setup to evaluate the matching estimators above.

There are 985 children in the dataset, of which 377 received the treatment of interest. We consider the same $d = 25$ covariates $X_i$ (6 continuous and 19 binary) used by [13] and normalize these in the same manner. The covariates include physical measurements of the child at birth such as weight, mother behavior during pregnancy such as smoking, and mother characteristics at time of birth such as marital status and education. We let $Y_i(0)$ be generated in the same way as the nonlinear response of [13] and using $\epsilon_{0i} \sim \mathcal{N}(0, 0.1)$.

Following [13], we prune the data to simulate an observational setting, but we consider a somewhat different pruning procedure. First, we sample $\beta_{\text{Treat}}$ uniformly at random from $\{-1, 1\}^d$ and assign the score $Q_i = \beta_{\text{Treat}}^T X_i + \|X_i\|_2 + \nu_i$ to each subject where $\nu_i \sim \mathcal{N}(0, 1)$ is a randomly and independently drawn standard normal random variable. Then, we prune away the half of the treated sample with the largest scores $Q_i$ and also prune away the half of the control sample with the smallest scores $Q_i$, leaving 492 subjects (with the same proportion of treated to control). Finally, we consider subsampling $n$ subjects at random from the pruned pool of 492 subjects.

For each $n = 10, 20, \ldots, 450$, we produce 100 replicates of the data, compute an estimate for SATT using each of the matching estimators listed above, and measure the RMSE over the 100 replicates. For CEM, the dimension of the data prohibits coarsening every covariate (coarsening each of the 25 covariates into only just two levels would result in over 33 million strata, and the probability that a stratum containing a treated unit wold also contain a control unit would be vanishing) and therefore we consider sampling just 3 dimensions at random and coarsening each into values above and below the mean. For PSM, we omit all replicates wherein the control and treated populations are perfectly separated by a hyperplane (necessarily, any sample with $n \leq d + 1$), which means that the logistic regression fit is undefined. We let $\sigma = \sqrt{d}/2 = 2.5$ for all kernels. We plot the results in Fig. 1(e).

Again, the results indicate a significant improvement due to kernel matching, which lead to RMSE that is nearly an order of magnitude smaller than most other methods across the board. The various non-linear kernel matching methods are very similar in performance, with the quadratic kernel slightly edging out the rest. The difference between matching with general probability weights or with a multisubset is nearly indistinguishable. Mean-matched sampling (linear kernel matching) performs less well than non-linear kernel matching but better than other methods, indicating a strong linear component that is still not prevalent enough to ignore the non-linear remanent.

Other estimators based on matching covariates, such as NNM and CEM, perform badly in this example due to the increased dimension of covariates. As the number of covariates increases, it becomes difficult to find units that are adequately similar on all dimensions, making the resulting one-to-one matching poor. The intuition extends to CEM, where it is impossible to exactly match on even very coarsely coarsened covariates due to their dimension, necessitating that we choose only a few covariates to match on, making the resulting match poor when considering all covariates. In comparison, matching the overall samples globally, as in kernel matching, instead of locally at the unit or coarsened stratum level, allows us to achieve much better balance while addressing estimation error directly. The failure of one-to-one matching to find good pairwise matches in the presence of moderate to high dimensions is cited by [33] as a reason to favor PSM. In this particular example, PSM (which is often not well-defined for $n \leq 100$) does better than one-to-one matching and CEM but worse than kernel matching (including linear kernel). The latter observation can be justified by noting that PSM simulates a control covariate sample drawn from the treated population, mimicking a completely randomized experiment [18], whereas matching on the covariates and doing so in a global manner mimics a well-balanced controlled experiment [17].

## 4 Conclusion

We presented a novel framework for matching estimators for causal inference from observational data. The framework is based on minimizing the dual norm of the error operator with respect to a space of possible conditional expectation functions. Many existing methods common in practice appear to fit this framework. We developed new, kernel-based estimators using the framework and showed they satisfy consistency. Our new estimators prove exceedingly successful in comparative empirical studies of matching estimators.

# Supplementary Material

**Proofs**

*Proof of Theorem 1.* Let us write SATT as
$$\text{SATT} = \tfrac{1}{n_1}\sum_{i\in\mathcal{T}_1} Y_i - \tfrac{1}{n_1}\sum_{i\in\mathcal{T}_1} Y_i(0).$$
It is then clear that SATT differs from $\hat\tau_W$ only in the second term, that is,
$$\hat\tau - \text{SATT} = \tfrac{1}{n_1}\sum_{i\in\mathcal{T}_1} Y_i(0) - \sum_{i\in\mathcal{T}_0} W_i Y_i(0)$$
$$= \sum_{i=1}^n (-1)^{T_i+1} W_i Y_i(0)$$
$$= \sum_{i=1}^n (-1)^{T_i+1} W_i f_0(X_i) + \sum_{i=1}^n (-1)^{T_i+1} W_i \epsilon_i,$$
where we recognize the last term as $E_W$. For each term of $E_W$ we have
$$\mathbb{E}[(-1)^{T_i+1} W_i \epsilon_i \mid X, T]$$
$$= (-1)^{T_i+1} W_i \left(\mathbb{E}[Y_i(0)|X,T] - f_0(X_i)\right)$$
$$= (-1)^{T_i+1} W_i \left(\mathbb{E}[Y_i(0)|X] - f_0(X_i)\right) = 0,$$
where the first equality is by definition of $\epsilon_i$ and the fact that $W_i = W_i(X, T)$ and the second is by Assumption 1. $\square$

*Proof of Theorem 2.* Let $D$ be the distance matrix $D_{ii'} = \delta(X_i, X_{i'})$. For this choice of $(\mathcal{F}, \|\cdot\|)$, by linear optimization duality we get
$$\mathfrak{E}(W;\mathcal{F}) = \tfrac{1}{n_1} \sup_{v_i - v_{i'} \leq D_{ii'}\forall i,i'} \left(\sum_{i\in\mathcal{T}_1} v_i - \sum_{i\in\mathcal{T}_0} n_1 W_i v_i\right)$$
$$= \tfrac{1}{n_1}\min_S \quad \sum_{i,i'} D_{ii'} S_{ii'}$$
$$\text{s.t.} \quad S \in \mathbb{R}_+^{n\times n}$$
$$\sum_{i'=1}^n (S_{ii'} - S_{i'i}) = 1 \quad \forall i \in \mathcal{T}_1$$
$$\sum_{i'=1}^n (S_{ii'} - S_{i'i}) = -n_1 W_i \quad \forall i \in \mathcal{T}_0.$$
This describes a min-cost network flow problem with sources $\mathcal{T}_1$ with inputs 1, sinks $\mathcal{T}_0$ with outputs $W_i$, edges between every two nodes with costs $D_{ii'}$ and without capacities. Consider any source $i \in \mathcal{T}_1$ and any sink $i' \in \mathcal{T}_0$ and any path $i, i_1, \ldots, i_m, i'$. By the triangle inequality, $D_{ii'} \leq D_{ii_1} + D_{i_1 i_2} + \cdots + D_{i_m i'}$. Therefore, as there are no capacities, it is always preferable to send the flow from the sources to the sinks along the direct edges from $\mathcal{T}_1$ to $\mathcal{T}_0$. That is, we can eliminate all other edges and write
$$\mathfrak{E}(W;\mathcal{F}) = \tfrac{1}{n_1}\min_S \quad \sum_{i\in\mathcal{T}_1, i'\in\mathcal{T}_0} D_{ii'} S_{ii'}$$
$$\text{s.t.} \quad S \in \mathbb{R}_+^{\mathcal{T}_1\times\mathcal{T}_0}$$
$$\sum_{i'\in\mathcal{T}_0} S_{ii'} = 1 \quad \forall i \in \mathcal{T}_1$$
$$\sum_{i\in\mathcal{T}_1} S_{ii'} = n_1 W_i \quad \forall i' \in \mathcal{T}_0.$$

In the case of with replacement and $\mathcal{W}_0 = \mathcal{W}_0^{\text{probability}}$, using the transformation $W_i' = n_1 W_i$, we get
$$\min_{W\in\mathcal{W}} \mathfrak{E}(W;\mathcal{F})$$
$$= \tfrac{1}{n_1}\min_{S,W'} \quad \sum_{i\in\mathcal{T}_1, i'\in\mathcal{T}_0} D_{ii'} S_{ii'}$$
$$\text{s.t.} \quad S \in \mathbb{R}_+^{\mathcal{T}_1\times\mathcal{T}_0}$$
$$W_i' \in \mathbb{R}_+^{\mathcal{T}_0}$$
$$\sum_{i\in\mathcal{T}_0} W_i' = n_1$$
$$\sum_{i'\in\mathcal{T}_0} S_{ii'} = 1 \quad \forall i \in \mathcal{T}_1$$
$$\sum_{i\in\mathcal{T}_1} S_{ii'} - W_i' = 0 \quad \forall i' \in \mathcal{T}_0.$$
This describes a min-cost netwrok flow problem with sources $\mathcal{T}_1$ with inputs 1; nodes $\mathcal{T}_0$ with 0 exogenous flow; one sink with output $n_1$; edges from each $i \in \mathcal{T}_1$ to each $i' \in \mathcal{T}_0$ with flow variable $S_{ii'}$, cost $D_{ii'}$, and without capacity; and edges from each $i \in \mathcal{T}_0$ to the sink with flow variable $W_i'$ and without cost or capacity. Because all data is integer, the optimal solution of $W' = n_1 W$ is integer [1]. Hence, since $\mathcal{W}_0^{n_1\text{-multisubset}} \subseteq \mathbb{Z}/n_1$, the solution is the same when we restrict to $\mathcal{W}_0 = \mathcal{W}_0^{n_1\text{-multisubset}}$. This solution (in terms of $W'$) is equal to sending the whole input 1 from each source in $\mathcal{T}_1$ to the node in $\mathcal{T}_0$ with smallest distance and from there routing this flow to the sink, which corresponds exactly to one-to-one matching with replacement.

In the case of no replacement and $\mathcal{W}_0 = \mathcal{W}_0^{n_1^{-1}\text{-bounded}}$, using the transformation $W_i' = n_1 W_i$, we get
$$\min_{W\in\mathcal{W}} \mathfrak{E}(W;\mathcal{F})$$
$$= \tfrac{1}{n_1}\min_{S,W'} \quad \sum_{i\in\mathcal{T}_1, i'\in\mathcal{T}_0} D_{ii'} S_{ii'}$$
$$\text{s.t.} \quad S \in \mathbb{R}_+^{\mathcal{T}_1\times\mathcal{T}_0}$$
$$W_i' \in \mathbb{R}_+^{\mathcal{T}_0}$$
$$\sum_{i\in\mathcal{T}_0} W_i' = n_1$$
$$W_i' \leq 1 \quad \forall i \in \mathcal{T}_0$$
$$\sum_{i'\in\mathcal{T}_0} S_{ii'} = 1 \quad \forall i \in \mathcal{T}_1$$
$$\sum_{i\in\mathcal{T}_1} S_{ii'} - W_i' = 0 \quad \forall i' \in \mathcal{T}_0.$$
This describes the same min-cost netwrok flow problem except that the edges from each $i \in \mathcal{T}_0$ to the sink have a capacity of 1. Because all data is integer, the optimal solution of $S$ and $W' = n_1 W$ is integer [1]. Hence, since $\mathcal{W}_0^{n_1\text{-subset}} \subseteq \mathbb{Z}/n_1$, the solution is the same when we restrict to $\mathcal{W}_0 = \mathcal{W}_0^{n_1\text{-subset}}$. The optimal $S_{ii'}$ is integer and so, by $\sum_{i'\in\mathcal{T}_0} S_{ii'} = 1$, for each $i \in \mathcal{T}_1$ there is exactly one $i' \in \mathcal{T}_0$ with $S_{ii'} = 1$ and all others are zero. $S_{ii'} = 1$ denotes matching $i$ with $i'$. The optimal $W_i'$ is integral and so, by $W_i' \leq 1$, $W_i' \in \{0,1\}$. Hence, for each $i \in \mathcal{T}_0$, $\sum_{i' \in \mathcal{T}_1} S_{ii'} \in \{0,1\}$ so we only use node $i$ at most once. The cost of $S$ is exactly the sum of pairwise distances in the match. Hence, the optimal solution corresponds exactly to one-to-one matching without replacement. □

*Proof of Corollary 3.* Apply Theorem 2 with the metric
$$\delta'(x, x') = \mathbb{I}_{[x \neq x']} \max\{\delta(x, x'), \delta_0\}. \quad \square$$

*Proof of Theorem 4.* This choice of space leads to
$$\mathfrak{E}(W; \mathcal{F}) = \sum_{j=1}^{M} \left| \frac{1}{n_1} \sum_{i \in \mathcal{T}_1} \mathbb{I}_{[C(X_i)=j]} - \sum_{i \in \mathcal{T}_0} W_i \mathbb{I}_{[C(X_i)=j]} \right|.$$

That is, the worst-case $f$ assigns $\pm 1$ to each partition in order to make the difference of values in that partition be nonnegative. Then clearly the optimal choice of $W \in \mathbb{R}^{\mathcal{T}_0}$ is to make each of these absolute values equal zero. This happens exactly when, for each $i \in \mathcal{T}_0$,

$$W_i = \frac{1}{n_1} \frac{|i' \in \mathcal{T}_1 : C(X_{i'}) = C(X_i)|}{|i' \in \mathcal{T}_0 : C(X_{i'}) = C(X_i)|}$$
$$= \frac{1}{n_1} \frac{\text{num treatment subjects in same partition as } i}{\text{num control subjects in same partition as } i},$$

where $0/0 = 0$ and we never encounter dividing a positive integer by 0 due to the no-extrapolation assumption. Because the weight is nonnegative, the solution is unchanged when restricting to nonnegative weights. □

*Proof of Theorem 5.* By duality of norms,
$$\mathfrak{E}(W; \mathcal{F}) = \sup_{\beta^T V \beta \leq 1} \beta^T \left( \frac{1}{n_1} \sum_{i \in \mathcal{T}_1} X_i - \sum_{i \in \mathcal{T}_0} W_i X_i \right)$$
$$= M_V(W).$$

The optimal $W$ minimizes this discrepancy over subsamples from control with the allowable size. □

*Proof of Theorem 6.* We have
$$\mathfrak{E}^2(W; \mathcal{F}) = \max_{\|f\| \leq 1} \left( \sum_{i=1}^{n} (-1)^{T_i+1} W_i f(X_i) \right)^2$$
$$= \left\langle \sum_{i=1}^{n} (-1)^{T_i+1} W_i \mathcal{K}(X_i, \cdot), \sum_{i=1}^{n} (-1)^{T_i+1} W_i \mathcal{K}(X_i, \cdot) \right\rangle$$
$$= \sum_{i,j=1}^{n} (-1)^{T_i+T_j} W_i W_j K_{ij},$$

which when written in block form gives rise to the result. □

*Proof of Theorem 7.* First we show $\mathfrak{E}_{\min}(\mathcal{F}) \to 0$ a.s. by showing that we can construct a feasible $\tilde{W}$ such that $\mathfrak{E}(\tilde{W}; \mathcal{F}) \to 0$ a.s. Let $p(x) = \mathbb{P}(T=1|X=x)$. By Assumption 1, $0 < p(X) < 1$ a.s. So there exists $\alpha > 0$ such that $q(x) = \alpha p(x)/(1-p(x))$ is a.s. in $(0,1)$. For each $i$, let $\tilde{W}'_i \in \{0,1\}$ be Bernoulli with probability $q(X_i)$. Then we have that $X_i|T=0, \tilde{W}'_i = 1$ is distributed as $X_i|T=1$. Let $n'_0 = \sum_{j \in \mathcal{T}_0} \tilde{W}'_j$ and note that $n'_0 \geq \underline{n}'_0$ eventually a.s. For each $i \in \mathcal{T}_0$, set $\tilde{W}_i = \tilde{W}'_i/n'_0$. Let $\zeta(f) = \mathbb{E}[f(X_1) | T=1]$, $\xi_i(f) = (T_i + \tilde{W}'_i)(f(X_i) - \zeta(f))$. Let $A_0 = \frac{1}{n_0} \sum_{i \in \mathcal{T}_0} \xi_i$ and $A_1 = \frac{1}{n_1} \sum_{i \in \mathcal{T}_1} \xi_i$. Adding and subtracting $\zeta$, we see $\mathcal{E}(\tilde{W}; f) = A_1(f) - (n_0/n'_0) A_0(f)$. By construction of $\tilde{W}'_i$, we see that $\mathbb{E}[\xi_i] = 0$ (i.e., Bochner integral). By (5), $\|\xi\|_*$ has (a) second or (b) first moment. By (1), each $\xi_i$ is independent. Therefore, by [2] for (5)(a) (since $B$-convexity of $\mathcal{F}$ implies $B$-convexity of $\mathcal{F}^*$; see [20]) or by [6] for (5)(b), a law of large numbers holds yielding, a.s., $\|A_0\|_* \to 0$ and $\|A_1\|_* \to 0$. Since $(n_0/n'_0) \to \alpha \mathbb{E}[p(X_1)] < \infty$ a.s., we have that $\|\mathcal{E}(\tilde{W}; \cdot)\|_* \to 0$ a.s. By (3) $\tilde{W}$ is feasible, so, a.s.

$$\mathfrak{E}_{\min}(\mathcal{F}) = \mathfrak{E}(W; \mathcal{F}) \leq \mathfrak{E}(\tilde{W}; \mathcal{F}) = \|\mathcal{E}(\tilde{W}; \cdot)\|_* \to 0.$$

Fix $\epsilon > 0$. By (4), theres is a $g_0 \in \mathcal{F}$ such that $\sup_x |f_0(x) - g_0(x)| \leq \epsilon/2$. Hence,

$$|\mathcal{E}(W; f_0)| \leq |\mathcal{E}(W; g_0)| + 2 \sup_{i=1,\ldots,n} |f_0(X_i) - g_0(X_i)|$$
$$\leq \|g_0\| \mathfrak{E}(W; \mathcal{F}) + \epsilon = \|g_0\| \mathfrak{E}_{\min} + \epsilon \to \epsilon.$$

Since true for any $\epsilon > 0$, $|\mathcal{E}(W; f_0)| \to 0$ a.s. □